\documentclass[sigconf]{acmart}

\copyrightyear{2023} 
\acmYear{2023} 
\setcopyright{acmcopyright}\acmConference[WSDM '23]{Proceedings of the Sixteenth ACM International Conference on Web Search and Data Mining}{February 27-March 3, 2023}{Singapore, Singapore}
\acmBooktitle{Proceedings of the Sixteenth ACM International Conference on Web Search and Data Mining (WSDM '23), February 27-March 3, 2023, Singapore, Singapore}
\acmPrice{15.00}
\acmDOI{10.1145/3539597.3570401}
\acmISBN{978-1-4503-9407-9/23/02}

\usepackage{subcaption}
\usepackage{hyperref}

\usepackage{soul}

\newcommand*{\sem}{\textsc{SEM}}
\newcommand*{\tsg}{\textsc{TwitterSG}}
\newcommand*{\birdwatch}{\textsc{BirdwatchSG}}

\AtBeginDocument{%
  \providecommand\BibTeX{{%
    \normalfont B\kern-0.5em{\scshape i\kern-0.25em b}\kern-0.8em\TeX}}}

\begin{document}

\title{Learning Stance Embeddings from Signed Social Graphs}

\author{John Pougu\'e-Biyong}
\affiliation{%
  \institution{University of Oxford}
  \city{Oxford}
  \country{UK}
}
\email{john.pougue-biyong@maths.ox.ac.uk}

\author{Akshay Gupta}
\affiliation{%
  \institution{Meta}
  \city{London}
  \country{UK}
}
\email{akshaykgupta@fb.com}

\author{Aria Haghighi}
\affiliation{%
  \institution{Twitter Cortex}
  \city{Seattle}
  \state{WA}
  \country{US}
}
\email{ahaghighi@twitter.com}

\author{Ahmed El-Kishky}
\affiliation{%
  \institution{Twitter Cortex}
  \city{Seattle}
   \state{WA}
  \country{US}
}
\email{aelkishky@twitter.com}
\renewcommand{\shortauthors}{Pougu\'e-Biyong, Gupta, Haghighi, El-Kishky}

\begin{abstract}
A challenge in social network analysis, is understanding the position, or stance, of people on a large set of topics. While past work has modeled (dis)agreement in social networks using signed graphs, these approaches have not modeled agreement patterns across a range of correlated topics. For instance, disagreement on one topic may make disagreement (or agreement) more likely for related topics. Recognizing topics influence agreement and disagreement, we propose the Stance Embeddings Model (\sem), which jointly learns embeddings for each user and topic in signed social graphs with distinct edge types for each topic. By jointly learning user and topic embeddings, \sem~can perform cold-start topic stance detection, predicting the stance of a user on topics for which we have not observed their engagement. We demonstrate the effectiveness of \sem\footnote{\label{note1}\url{https://github.com/lejohnyjohn/learning-stance-embeddings-from-signed-social-graphs}}~using two large-scale Twitter signed graph datasets that we open-source\footnote{\label{note2}\url{https://huggingface.co/datasets/Twitter/SignedGraphs}}. One dataset, \tsg, labels (dis)agreements using engagements between users via tweets to derive topic-informed, signed edges. The other, \birdwatch, leverages community reports on misinformation and misleading content. On \tsg~and \birdwatch, \sem~ shows a $39\%$ and $26\%$ error reduction respectively against strong topic-agnostic baselines.
\end{abstract}



\maketitle

\section{Introduction} %
Signed graphs (or networks) have been used to model support and opposition between members of a group of people, or community, in settings ranging from understanding political discourse in congress \cite{thomas2006} to identifying polarization in social networks \cite{leskovec2010signed, pougue2021debagreement}. In such graphs, each node represents an individual in the community, a positive ($+$) edge indicates agreement between two community members and a negative ($-$) one denotes disagreement. For instance, Epinions~\cite{leskovec2010signed} is a \emph{who-trust-whom graph} extracted from the now-defunct online review site, where each edge represents whether one member has rated another as trustworthy ($+$) or not  ($-$). The 108th US Senate signed graph~\cite{neal2014backbone} represents political alliances ($+$) or oppositions ($-$) between congressional members across 7,804 bills in the 108th U.S. Congress. Past work have leveraged signed graphs and insights from social psychology \cite{cartwright1956structural}  in order to better understand and predict patterns of community interaction \cite{leskovec2010signed, neal2014backbone}.

The intent of this work is to establish the value of leveraging topics in stance representations. Recent research in text-based stance detection has proven the benefits of capturing implicit relationships between topics, especially in cases where there are many topics at stake, and most with little training data~\cite{allaway2020zero, liu2021enhancing, allaway2021adversarial}. One shortcoming of traditional signed graph analysis is that it reduces the interaction between any two individuals to a binary value of agreement ($+$) or disagreement ($-$). Interactions in communities may be much more complex and change depending on underlying context. In the U.S. senate, two senators may agree on bills related to climate change, but differ on taxation policy bills. In a sports community, two French football fans may support rival clubs, but will generally both support the national team at the World Cup. Most communities will have several different aspects, or \textit{topics}, of discourse that have rich structure and dynamics within a community. For instance, in the French football fan example, it is very likely for someone to support the national team if we have observed support for a local club. This example and others highlight the value of modeling community stance across a range of topics~\cite{monti2021learning}.

In this work, we use \textit{signed topic graphs} to represent (dis)agreement across topics of discourse with a community. Each edge represents a binary agreement value ($\{+, -\}$) with respect to a single topic $t$; the inventory of topics is assumed to be fixed and finite, but varies across applications. Our proposed method, the Stance Embeddings Model (\sem), detailed in Section~\ref{sec:stance_embeddings}, leverages an extension of the node2vec algorithm \cite{grover2016node2vec} to signed topic graphs to learn embeddings for nodes as well as for topics. Learning member (node) and topic embeddings jointly enables us to represent topic-informed stance embeddings for each member, which can accurately predict member agreement across community topics (Section~\ref{predicting_link_sign}). This allows us to do zero-shot topic-stance prediction for a member, even when we haven't observed past engagement from the member on a topic (Section~\ref{subsec:zero-shot-stance-detection}). As importantly, it allows us to capture implicit relationships between topics (Section~\ref{subsec:visualisation}).

We apply and evaluate our approach on two Twitter-based signed social graphs that we open-source alongside this work (see Section~\ref{sec:datasets}). For both of these datasets, we represent online interactions as a signed topic graph, where each node is a Twitter user and each edge represents an interaction between users on a given topic. The \tsg~ dataset (Section~\ref{subsec:tsg}) consists of $\sim$13M interactions (edges) between $\sim$750k Twitter users (nodes), spanning 200 sports-related topics; each edge represents one user replying to another user's Tweet or explicitly using  the 'favorite' UI action (AKA, a \textit{like}). This graph is $\sim$6x larger than the Epinions graph, which to the best of our knowledge, is the largest publicly available signed social graph. The \birdwatch~ dataset instead leverages Birdwatch\footnote{\url{https://blog.twitter.com/en_us/topics/product/2021/introducing-birdwatch-a-community-based-approach-to-misinformation}} annotations to indicate whether a user finds information on a Tweet to be misinformation or misleading or are rated helpful in clarifying facts (see Section~\ref{subsec:birdwatch}~for details).

The core contributions of this paper are:
\begin{itemize}
    \item \textbf{Stance Embeddings Model (\sem)}: Generalisation of \newline node2vec to signed topic graphs. The model enables us to consider both topic and (dis)agreement for each edge during training, allowing to understand how topics relate to each other, how users engage with topics, and how users relate to each other across topics, even for topics a user may not have engaged with. In particular, we show that our topic-aware model improves on stance detection and learns topic relationships in an unsupervised fashion.
   \item \textbf{Datasets} Two signed topic graph datasets built with Twitter data, suitable for future research on understanding topical stance in large-scale communities and valuable resource for the graph mining community.
\end{itemize}
We open source our code and datasets on GitHub and HuggingFace to the community.

\section{Related work} %

Our work falls within the domain of shallow, signed and edge-attributed graph embeddings. Shallow graph embedding methods learn node embeddings when node features are unavailable~\cite{hamilton2020graph} by only utilizing the structure (i.e., the adjacency matrix) of a graph.

\subsection{Unsigned graph embeddings} 
\label{subsec:unsigned_graph_embeddings}
Most graph embedding techniques assume graphs are \textit{unsigned} -- that is edges indicate a single type of interaction (e.g., positive)~\cite{el2022graph}. Some popular methods include node2vec~\cite{grover2016node2vec}, deepwalk~\cite{perozzi2014deepwalk}, and LINE~\cite{tang2015line}. Node2vec and DeepWalk build on top of word2vec~\cite{mikolov2013efficient}, a word embedding technique in natural language processing. Node2vec generates second-order random walks on unsigned graphs, and learns node embeddings by training a skip-gram with negative sampling (SGNS)~\cite{mikolov2013distributed} to predict the surroundings of the input node. The learnt embeddings possess the property that nodes close in the graph are close in the embedding space. Deepwalk is a specific case of node2vec: it generates unbiaised random walks and follows the same subsequent process. \cite{qiu2018network} show that LINE implicitly is a special case of deepwalk. See~\cite{hamilton2020graph} for a more comprehensive survey. Such methods are not adapted for signed graphs because they are based on the homophily assumption (connected nodes lie close in the embedding space) whereby in signed graphs agreeing nodes should be closer while disagreeing nodes farther apart. Previous works on heterogeneous graph embeddings~\cite{elkishktwhin,dong2017x, wang2021embedding,el2022knn} also assume homophily.

\subsection{Signed graph embeddings}
\label{subsec:signed_graph_embeddings}

\textit{\hspace{2pt}\hspace{2pt} Spectral methods.\hspace{2pt}} The earliest approaches to learn signed graph embeddings are spectral. Spectral methods compute a low-rank matrix factorization of the signed graph Laplacian and study its spectrum~\cite{kunegis2010spectral, kunegis2014applications, zheng2015spectral, zheng2021modeling}.

\textit{Random walk-based methods.\hspace{2pt}} Other methods address the homophily limitation in signed graphs include random walk-based embeddings. SNE~\cite{yuan2017sne} generates uniform random walks on signed graphs (by ignoring the edge signs) and replaces the skip-gram model by a log-bilinear model. The model predicts the representation of a target node given its predecessors along a path. To capture the signed relationships between nodes, two signed-type vectors are incorporated into the log-bilinear model. SIDE~\cite{kim2018side} generates first-order random walks and defines a likelihood function composed of a signed proximity term to model the social balance theory, and two bias terms to mimic the preferential attachment theory. Similarly, SIGNet~\cite{islam2018signet} designs a new sampling technique for directed signed networks, and optimises an objective function that learns two representations for each node. BESIDE~\cite{chen2018bridge} mathematically models \textit{bridge} edges based on balance and status theory.

\textit{Deep learning-based methods.\hspace{2pt}} The signed graph embedding problem has also been tackled with deep learning-based approaches. SiNE~\cite{wang2017signed} is a deep neural network-based model that maximises the margin between the embedding similarity of friends and the embedding similarity of foes. StEM~\cite{rahaman2018method} is a deep learning method aiming at learning not only representations of nodes of different classes (e.g. friends and foes) but also decision boundaries between opposing groups . Hence, unlike other methods such as SiNE which are distance-based (using only local information), StEM attempts to incorporate global information. Instead of relying on social theories, ROSE~\cite{javari2020rose} learns role similarities between nodes.

\textit{Our model.\hspace{2pt}} Previous work on signed graph embeddings operates on edges while ignoring valuable information such edge attributes. Our work extends the traditional SGNS of word2vec and node2vec to signed graphs while the valuahle information found in edge attributes (e.g., topics). We do this by ensuring that each training example is constructed via a topic- and sign-informed random walk. Extending the skip-gram architecture not only provides scalability advantages, but also the flexibility to principally incorporate edge attributes. We demonstrate that incorporating edge attributes, such as topics, in the embedding process can benefit understanding stance in signed social-network interactions.
\section{Learning Stance Embeddings} \label{sec:stance_embeddings}

\subsection{Preliminaries}
\label{subsec:overview}
Let $G = (V, E)$ be a signed (un)directed topic graph: each edge has a topic $t$, and a sign of $-$ or $+$. We use $T$ to denote the finite set of topics $t$. Note that there can be multiple edges between users corresponding to different topic interactions. We define $G_t = (V_t, E_t)$ the subgraph of $G$ which contain all the edges with topic $t$.  We aim to learn a node mapping function $f_V : V \rightarrow \mathbb{R}^d$ , and a topic embedding function $f_T : T \rightarrow \mathbb{R}^d$.

Our approach will define embeddings for each edge using learned node and topic embeddings. For an edge $(u, v)$ with topic $t$, we combine the source embedding and topic embedding using $\sigma(f_V(u), f_T(t))$; see Section~\ref{models} for choices of $\sigma$ considered. This transformed source node embedding is combined with the target node embedding using an operator $\Phi(\cdot, \cdot)$ from Table~\ref{tab:operations}. We evaluate these edge embeddings compared to other signed graph edge embeddings in Section~\ref{sec:experiments}, but for the remainder of this section, we will detail how we learn the node and topic embedding functions $f_V$ and $f_T$. 

\subsection{Training data creation}
As we apply the skip-gram objective to graph data via random walks, our work can be considered an extension to node2vec~\cite{grover2016node2vec}. However, while node2vec only operates on unsigned homogeneous graphs, our embedding approach naturally incorporates signed edges as well as edge attributes such as topics.

Given an input signed topic graph, we outline how we create training examples to learn node and topic embeddings using the skip-gram objective.

\subsubsection*{Random walks on edge-attributed graphs}

We first iterate through each topic-specific subgraph $G_t$, and mask the edge weights yielding a topic-graph $G'_t = (V_t, E'_t)$ where all edges are unsigned and unweighted. We follow the sampling procedure of~\cite{grover2016node2vec}, and define a second-order random walk with two parameters $p$ and $q$ that guide the walker on $G'_t$. Let us consider a walker that just traversed edge $(s, u)$ and now resides at node $u$. The walker next decides to walk to edge $(u, v)$ with the unnormalised transition probability $\pi_{uv}$:

\begin{align} \label{eq:transition_probs}
\pi_{uv} = \left\{
    \begin{array}{ll}
        \frac{1}{p} & \mbox{if } v = s \\
        1 & \mbox{if } d_{sv} = 1 \\
        \frac{1}{q} & \mbox{if } d_{sv} = 2 \\
    \end{array}
\right.
\end{align}

where $d_{sv}$ is the shortest path distance between nodes $s$ and $v$. $p$ and $q$ are return and in-out parameters respectively, and control how fast the walk explores and leaves the neighborhood of starting node $s$. For example, $q < 1$, means the walker is more inclined to visit nodes which are further away from node $s$.

For each node $n$ in $G'_t$, we simulate $r$ random walks of fixed length $l$ starting at $n$. At every step of the walk, sampling is done based on transition probabilities defined in Eq.~\ref{eq:transition_probs}.

\subsubsection*{Creating signed contexts} In node2vec, the contexts of a source node $u$ are the nodes surrounding it in the walks. The context vocabulary $C$ is thus identical to the set of nodes $V$. This effectively embeds connected node close to each other in the embedding space. However, in signed graphs, agreeing nodes (linked with positive edges) should be embedded in close proximity while  disagreeing nodes (linked with negative edges) should be farther away. We incorporate these insights into our skip-gram objective.

Unlike with node2vec, whereby a source node predicts context node, we propose to predict \textit{sign and node} as contexts. In other words, we predict not only the context node, but also whether the source node agrees or disagrees with them on a given topic. While the context node is determined by the random walk, there may not be a signed edge between a source node and context node for that topic. To infer whether or not a source and context node agree on some topic,  we apply Heider's social balance theory~\cite{cartwright1956structural}.

\begin{figure}[h]
  \centering
  \includegraphics[width=0.8\linewidth]{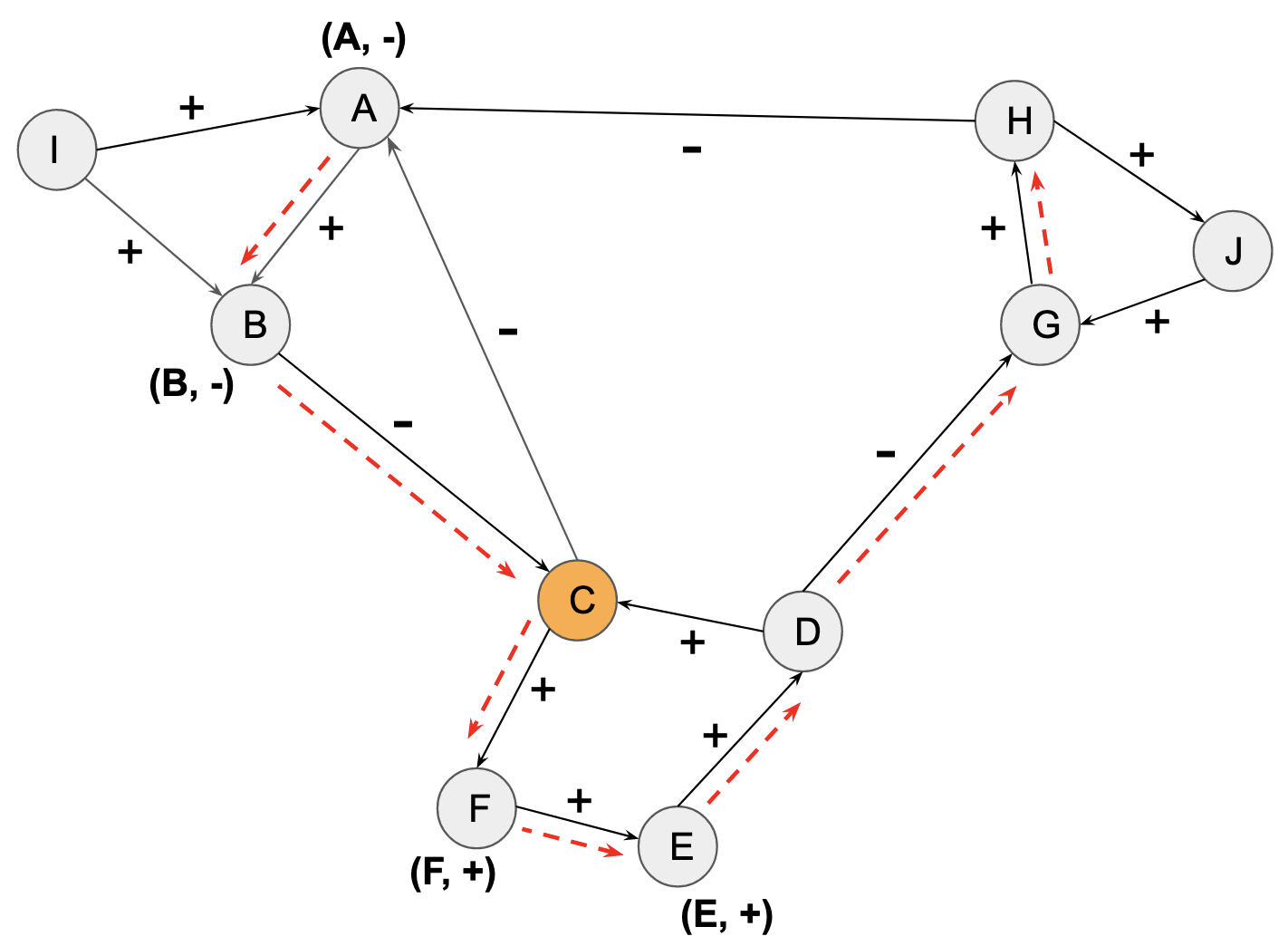}
  \caption{A sample random walk (in red) on a signed graph. The corresponding sign-informed contexts for source node $u_o=C$ are shown in bold (assuming a window of size 2).}
  \Description{A signed graph with a sample random walk and a sign-informed context example.}
  \label{fig:sign_info}
\end{figure}

Let $t$ be an arbitrary topic, and consider the graph $G_t$ depicted in Figure \ref{fig:sign_info}. Assuming a random walk sampled via the procedure described above, we have a sequence of nodes. Using a window of size $k$ around a source node $u_0$, $2k$ context nodes are produced from the walk: $k$ before $u_0$ and $k$ after: $(u_{-k} \ldots u_0 \ldots u_{+k})$. In addition we compute the inferred sign, $S(u_0, u_i)$, between our source node and the $i_{th}$ context node as follows: 

\begin{align} \label{eq:context_sign}
S(u_0, u_i) = \left\{
    \begin{array}{ll}
       \prod_{m=i+1}^{0} w_{u_{m-1}u_{m}} & i < 0 \\[1em]
        \prod_{m=1}^{i} w_{u_{m-1}u_{m}} & i > 0 \\
    \end{array}
\right.
\end{align}
where $w_{uv}$ is the weight, $+1$ or $-1$, between nodes $u$ and $v$.

As seen in Equation~\ref{eq:context_sign}, we can leverage Heider's social balance theory to assign each context node a sign with respect to the source node. In simple terms we have three rules: (i) the friend ($+$) of my friend ($+$) is my friend ($+$), (ii) the friend ($+$) of my enemy ($-$) is my enemy ($-$), and (iii) the enemy ($-$) of my enemy ($-$) is my friend ($+$). Equation~\ref{eq:context_sign} applies this to (dis)agreements over topics and as such, we can compute the (dis)agreement sign between the source node and a context node by multiplying the edge signs between the source and context as defined by the random walk between them.%

By incorporating these (dis)agreements with the source node alongside each context node, our skip-gram objective needs to predict both the context node and its agreement on a topic. As such, node proximity and stance both influence a node's embedding. We apply social balance theory on a per-topic network basis as we do not have predefined knowledge about topic associations in general. However, we show in section~\ref{subsec:visualisation} and Figure~\ref{fig:tsg_topics} that these associations can automatically be learnt by our model during training.

\subsection{Learning node \& topic embeddings}
\label{subsec:joint_learn}

\begin{figure}[h]
  \centering
  \includegraphics[width=0.9\linewidth]{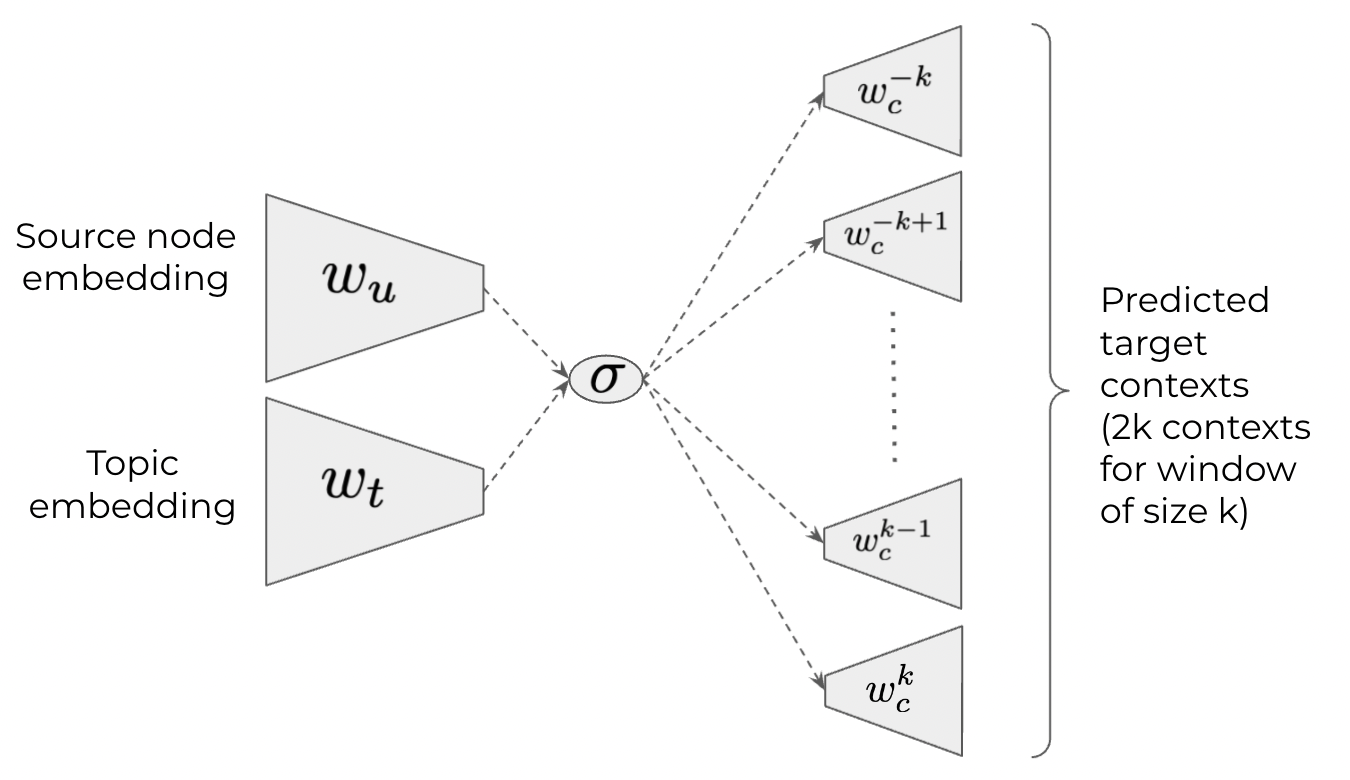}
  \caption{Skip-gram architecture for \sem takes source node and topic as input and predicts sign-aware contexts.}
  \Description{Our proposed skip-gram architecture. It combines source node and topic embeddings to predict sign-informed contexts.}
  \label{fig:skipgrams}
\end{figure}

The training examples are composed of a source node $u$, a topic $t$, and a set of contexts $C_{t}(u)$ where contexts consist of \textit{(node, sign)} pairs. We associate embedding vectors $W_u$, $W_c$, and $W_t$ for the source, context (node-sign pair), and topics respectively; these vectors are parameters to be learned. In Fig.~\ref{fig:skipgrams}, we visualize this  topic-aware skip-gram architecture as a generalisation of the original skip-gram neural network architecture.

To learn these vectors, we generalise the SkipGram objective to incorporate topic information $t$ as follows: 

\begin{align} \label{eq:newobjective}
    \max_{W} \quad & \sum_{t \in T} \sum_{u \in V} \left[  -\log Z_{u, t} + \sum_{c \in C_{ t}(u)} W_c \cdot \sigma(W_t, W_u) \right]
\end{align}

\noindent where $Z_{u ,t}= \sum_{c' \in C_t} \exp(W_{c'} \cdot \sigma(W_t, W_u))$, with $\sigma(\cdot, \cdot)$ an operation over topic and node embedding vectors (e.g. addition of both vectors).  The sign in any context $c$ of Equation~\ref{eq:newobjective} is derived from Equation~\ref{eq:context_sign}. The dot product $W_c \cdot \sigma(W_t, W_u)$, with $c=(v,\pm)$, measures the similarity between user $u$'s topic-aware embedding $\sigma(W_t, W_u)$ and user $v$'s perspective embedding $W_{c}$. As the partition function $Z_{u,t}$ is expensive to compute, we approximate it using negative sampling~\cite{mikolov2013distributed}. Note that we could somehow tie parameterisation between the contexts $(u, +)$ and $(u, -)$ but we choose not to as past skip-gram work has found allocating distinct context vectors for the same underlying entity to be effective (see, e.g, ~\cite{song-etal-2018-directional}). 

\section{Datasets} %
\label{sec:datasets}
To evaluate our approach we curate two new social-network signed topic graphs that we open-source alongside our work. Both datasets are fully anonymized without personally identifiable information.

\subsection{\tsg} %
\label{subsec:tsg}
Twitter Signed Graph, or \tsg, is a signed, directed, edge-attributed graph of users, drawn from Twitter interactions. A positive-signed edge exists from user $A$ to user $B$ if user $A$ liked a tweet posted by user $B$. A negative-signed edge exists from user $A$ to user $B$ if user $A$ expressed opposition towards user $B$'s tweet, e.g., by replying \textit{I disagree with you}. The topic of an edge from user $A$ to user $B$ is determined by the topic of user $B$'s tweet, also called the \textit{target tweet}. Tweet topics were inferred with a proprietary tweet topic classifier used in production at Twitter; we restrict interactions in \tsg~to sports-related topics (e.g., sports teams, players, managers, or events). The tweets related to these interactions were published between 20th May (Ice Hockey World Championships) and 8th August 2021 (closing date of the 2020 Tokyo Olympic Games).

Several challenges arise when attempting to build a large signed graph with interactions on Twitter. First, the graph may be extremely sparse due to the number of active users and the skewed distribution of tweets per user. Second, opposition mostly goes silent (the user may keep scrolling if they do not agree with a statement) or is expressed via reply to a tweet, which requires more effort than clicking a \textit{like} button to express support. As such, there is an unbalance between the amount of support and opposition signals. And lastly, opposition in a tweet may be implicit.

\begin{figure}[h]
  \centering
  \includegraphics[width=1.0\linewidth]{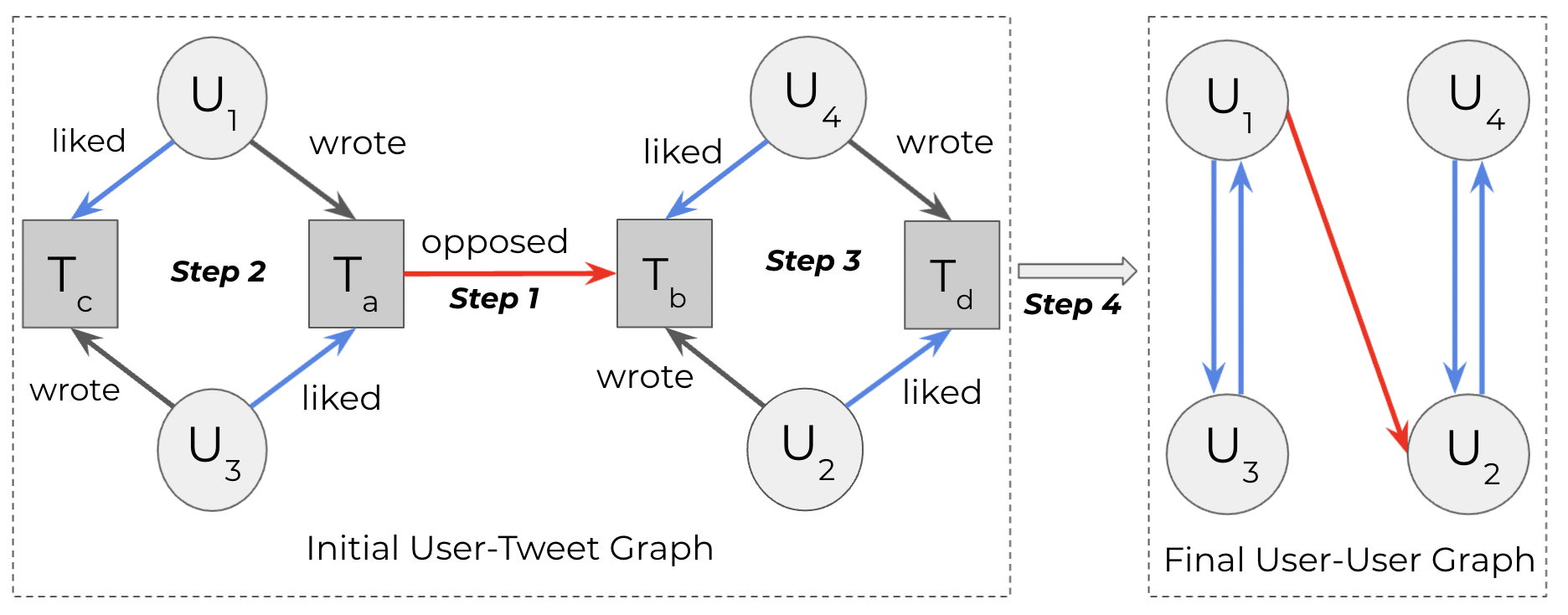}
  \caption{Steps involved in building \tsg. The final user-user graph is obtained following step 4 of Section~\ref{subsec:tsg}.}
  \Description{A diagram depicting the steps to build TwitterSG from user-tweet interactions.}
  \label{fig:building_twitter_SG}
\end{figure}

To overcome these challenges, we first create a user-tweet graph (Fig. \ref{fig:building_twitter_SG}) that we project onto a user-user graph:
\begin{enumerate} 
    \item For the sake of simplicity, we curated a list of high-precision English and French expressions which express clear opposition (e.g. ``I disagree" and ``you're wrong"). We retained all sports-related tweets $T_a$ containing at least one of these expressions, and the tweets $T_b$ they replied to. For the sake of clarity, tweet $T_a$ ($T_b$) is posted by user $U_1$ ($U_2$).
    \item To control the graph sparsity, we retained all users $U3$ who both (i) wrote a tweet $T_c$ liked by user $U1$, and (ii) liked the tweet $T_a$ (opposition tweet) written by user $U1$. 
    \item Similarly, we retained all users $U4$ who both (i) wrote a tweet $T_d$ liked by user $U2$, and (ii) liked the tweet $T_b$ written by user $U2$. At this stage, the positive interactions largely outnumbered the negative ones (300k negative interactions for more than 100M positive ones). Filtering out a large portion of positive edges would increase the share of negative edges but would decrease the number of users. Conversely, filtering in a large portion of positive edges would push the share of negative edges close to 0 but increase the number of users. We found a trade-off cut by selecting a share of likes (retrieved in steps (2) and (3)) so that the share of negative edges in our graph is close to 10\%. We ranked the topics by decreasing frequency and filtered out all the tweets not related to the top 200 topics. 
    \item We project the resulting user-tweet graph onto a user-user graph. We anonymise all the nodes (users) and edges (tweets).
    \end{enumerate}
The edge data of the final graph is provided under a simple 4-column table format: \textit{source node}, \textit{target node}, \textit{topic}, \textit{edge weight} ($\pm1$).

\tsg~ contains 753,944 nodes (users), 200 topics and 12,848,093 edges. Among these edges, 9.6\% are negative (opposition) and 90.4\% are positive. 
\begin{figure}[h]
  \centering
  \includegraphics[width=1.0\linewidth]{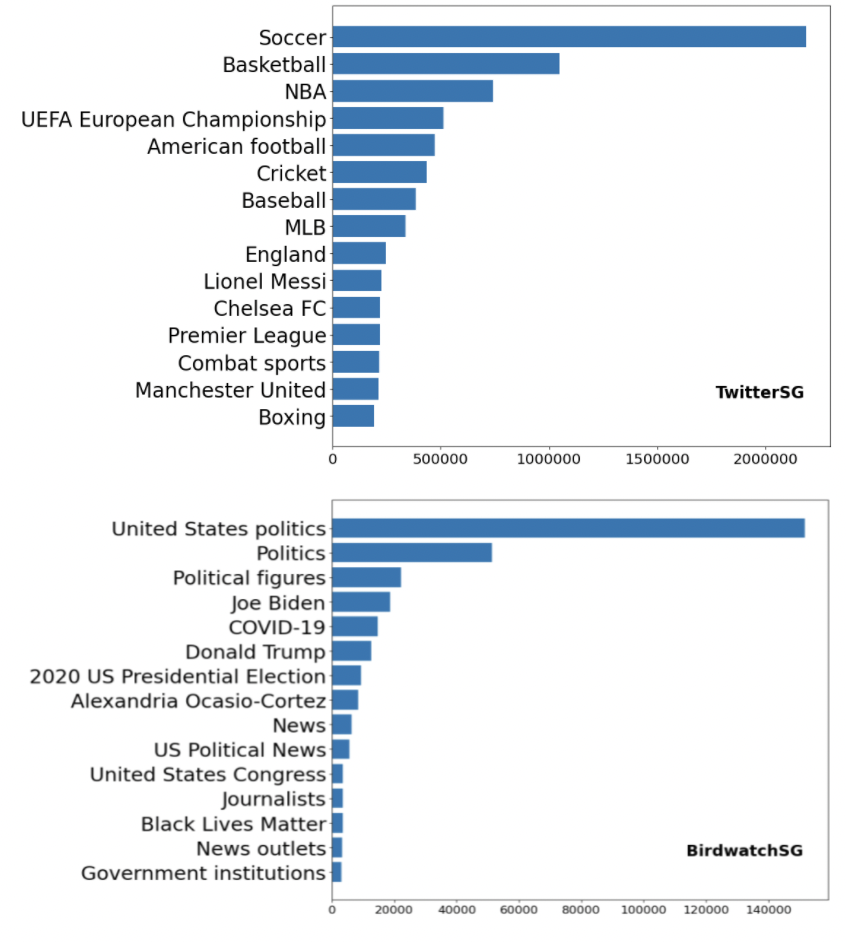}
  \caption{Top-15 topics in \tsg~(top) and \birdwatch~(bottom), ordered by decreasing frequency.}
  \Description{Bar histogram of the Top-15 topic frequency in BirdwatchSG.}
  \label{fig:topic_distribution_stacked}
\end{figure}
Most frequent topics are depicted in Figure~\ref{fig:topic_distribution_stacked}.

\subsection{\birdwatch} %
\label{subsec:birdwatch}
Birdwatch Signed Graph,  or \birdwatch,  is  a  signed,  directed,  edge-attributed graph of users, drawn from note ratings on Birdwatch\footnote{\url{https://blog.twitter.com/en_us/topics/product/2021/introducing-birdwatch-a-community-based-approach-to-misinformation}}. Birdwatch is a pilot launched by Twitter in January 2021 in the USA to address misleading information on the platform, in a community-driven fashion: the Birdwatch participants can identify information in tweets they believe is misleading and write notes that provide informative context. They can also rate the helpfulness (either \textit{helpful}, \textit{somewhat helpful}, or \textit{not helpful}) of notes added by other contributors. 
All Birdwatch contributions are publicly available on the Download Data page of the Birdwatch site\footnote{\url{https://twitter.github.io/birdwatch/}} so that anyone in the USA has free access to analyse the data.

Starting with Birdwatch data from January to July 2021, we create a positive (negative) edge from participant $U_1$ to $U_2$ if participant $U_1$ rated a note written by participant $U_2$ as \textit{helpful} (\textit{not helpful}). We filter out the \textit{somewhat helpful} ratings. The topic associated with an edge is the topic of the tweet the note refers to. We anonymise all the nodes and edges. The edge data of the final graph is provided under a 4-column format similar to \tsg~(Section \ref{subsec:tsg}).

The graph contains 2,987 nodes (users), 1,020 topics and 441,986 edges. Among these edges, 36.9\% are negative (opposition) and 63.1\% are positive. Most frequent topics are depicted in Figure~\ref{fig:topic_distribution_stacked}. There may be several edges between two nodes (several topics, several interactions).

\subsection{Comparison with existing signed graphs} 

Epinions, Slashdot~\cite{leskovec2010signed}, Wiki-Rfa~\cite{west2014exploiting}, BitcoinOtc, BitcoinAlpha~\cite{kumar2016edge, kumar2018rev2} are the largest and most widely used signed social graphs for benchmarking signed graph embeddings methods. Epinions.com\footnote{\url{https://en.wikipedia.org/wiki/Epinions}} was a product review site where users can write reviews for various products with rating scores from 1 to 5. Other users could rate the helpfulness of reviews. Slashdot\footnote{\url{https://slashdot.org/}} is a technology news platform on which users can create friend and foe links with other users. For a Wikipedia editor to become an administrator, a request for adminship (RfA) must be submitted, and any Wikipedia member may cast a supporting, neutral, or opposing vote. This induces a directed, signed graph Wiki-Rfa~\cite{west2014exploiting} in which nodes represent Wikipedia members and edges represent votes. BitcoinOtc and BitcoinAlpha~\cite{kumar2016edge, kumar2018rev2} are who-trusts-whom graphs of users who trade using Bitcoin on online platforms. Since Bitcoin users are anonymous, there is a need to maintain a record of users' reputation to prevent transactions with fraudulent and risky users. Platforms’ members can rate each  other members positively or negatively.

\paragraph{Our datasets} Our two real-world signed social graphs contain attributed (topics) edges. \tsg~consists of $\sim$13M interactions (edges) between $\sim$750k Twitter users (nodes), spanning 200 sports-related topics: teams, sports, players, managers, and events (e.g. Los Angeles Lakers, Basketball, Olympics). It contains $\sim$6x more nodes than Epinions, the largest publicly available signed graph. \birdwatch~consists of $\sim$3k Birdwatch participants, $\sim$450k edges spanning 1,020 diverse topics prone to misleading content and/or partisanship (e.g. COVID-19, US Presidential Elections). Table \ref{tab:signed_graphs_statistics} provides a comparison of all the datasets.
\begin{table}
  \caption{Statistics of signed graph datasets. The bottom two denote datasets released as part of this work.}
  \label{tab:signed_graphs_statistics}
  \begin{tabular}{cccl}
    \toprule
    Dataset & $|V|$ & $|E|$ & $\%_{|E_{-}|}$ \\
    \midrule
    BitcoinAlpha & 3,783 & 24,186 & 7\% \\
    BitcoinOtc & 5,881 & 35,592 & 9\% \\
    Epinions & 131,828 & 841,372 & 15\% \\
    Slashdot & 77,357 & 516,575 & 23\% \\
    Wiki-Rfa & 10,835 & 159,388 & 22\% \\
    \midrule
    \birdwatch & 2,987 & 441,986 & 37\% \\
    \tsg & 753,944 & 12,848,093 & 10\% \\
  \bottomrule
\end{tabular}
\end{table} 

\section{Experiments}
\label{sec:experiments}
In this section, we demonstrate the value of topic information to understand stances. To explore this, we evaluate the embeddings produced by our \sem~method (Section~\ref{sec:stance_embeddings}) and compare its performance to three topic-agnostic state-of-the art signed graph embedding models on our \tsg~and \birdwatch~datasets (Section~\ref{sec:datasets}). 

\subsection{Embedding Models}
\label{models}
\paragraph{\sem~variants}
We evaluate three variants of \sem, each of which corresponds to a different choice of $\sigma$ function to combine node and topic embeddings (Section~\ref{sec:stance_embeddings}):
\begin{itemize}
    \item \sem-\textit{mask}: The topic information is ignored. This corresponds to $\sigma(W_t, W_u) = W_u$ in the first layer of the topic-aware skip-gram architecture, Fig. \ref{fig:skipgrams}.
    \item \sem-\textit{addition}: The topic and node embeddings are added in the first layer of the topic-aware skip-gram architecture (Fig. \ref{fig:skipgrams}), i.e., $\sigma(W_t, W_u) = W_t + W_u$. 
    \item \sem-\textit{hadamard}: The topic and node embeddings are combined via element-wise multiplication (hadamard) in the first layer of the topic-aware skip-gram architecture, i.e., $\sigma(W_t, W_u) = W_t \times W_u$.
\end{itemize}

Note that the \sem~variants only change how the user and topic embedding are combined during skipgram training (Section~\ref{subsec:joint_learn}).

\paragraph{Baselines} We compare \sem~to three state-of-the-art signed graph embedding methods described in Section~\ref{subsec:signed_graph_embeddings}: StEM~\cite{rahaman2018method}, SIDE~\cite{kim2018side}, SiNE~\cite{wang2017signed}. We chose these methods based on their competitive performance, code availability and reproducibility. Like \sem-\textit{mask}, these three methods are topic agnostic and were only tested on signed graphs lacking topics, or other attributes, on edges.

\subsection{Training setup}
We set the node embedding dimension ($d$) to 64 for all methods and experiments\footnote{We experimented with a standard embedding size, however the effects of embedding size on performance should mimic those displayed by node2vec~\cite{grover2016node2vec} as \sem~is a topic and sign-aware extension of it.}. For \sem~variants, we set walks per node $r \in \{5, 10, 20, 80\}$, walk length $l=40$, context size $k=5$, return parameter $p=1.5$, in-out parameter $q=0.5$, negative sample size to $20$, subsampling threshold to 1e-5, and the optimisation is run for 1 to 5 epochs. For two given users and a given topic, edge weights are summed and the overall topical edge weight is set to +1 if the sum is positive, and -1 otherwise. For baseline methods, we use the same parameter settings as those suggested in their respective papers. The edge topic information is masked for baselines and \sem-\textit{mask}.

\subsection{Evaluation setup} \label{evaluation_setup}
We follow previous work by evaluating our method, \sem, and baselines on a signed link prediction task~\cite{rahaman2018method, kim2018side, wang2017signed}. In signed link prediction, we are given a signed graph where the sign, or agreement value, on several edges is missing and we predict each 
edge's sign value using the observed edges. In particular, we formulate link sign prediction as a binary classification task using embedding learned from each method as follows.  For each dataset, we perform 5-fold cross-validation (80/20\% training/test set) and evaluate with mean AUC over the 5 folds. For all approaches, we create edge embeddings by combining node embeddings using $\Phi(u_1, u_2)$ using operations from Table~\ref{tab:operations}. Note that this means for topic-aware \sem~variants we do not explicitly use the topic embedding for evaluation. 
\begin{table}[h]
  \caption{Operations ($\Phi$) to produce edge embeddings from node embeddings for evaluation ( Section~\ref{evaluation_setup})}
  \label{tab:operations}
  \begin{tabular}{ c  l}
    \toprule
     Operation  & Output \\
    \midrule
    hadamard & $ w[i] = u_1[i] \times u_2[i]$  \\
    $\ell_1$ &  $w[i] = | u_1[i] - u_2[i] |$ \\
    $\ell_2$ &  $w[i] = (u_1[i] - u_2[i])^2$ \\
    Average & $w[i] =\frac{1}{2}(u_1[i] + u_2[i])$  \\
    Concatenation &  $w = u_1 \otimes u_2$  \\
  \bottomrule
\end{tabular}
\end{table}
Using the edge representations in the training set, we fit a binary classifier to predict edge signs on the test set. Due to the sign imbalance sign in the edge data, we downsample the positive signs when fitting the classifier.

\subsection{Stance detection: predicting link sign} \label{predicting_link_sign}

In Table~\ref{tab:simple_knn_lr}, we report results for \sem~variants and baselines using both nearest neighbors (kNN) and logistic regression (LR) classification on edge embeddings. For each approach, we report the best value over choices of translation operator $\Phi(\cdot, \cdot)$ from Table~\ref{tab:operations}.
\begin{table}[h]
\caption{Mean AUC from 5-fold CV on stance detection using nearest neighbors (kNN) and logistic regression (LR) on edge embbedings to predict stance (Section~\ref{evaluation_setup}).} 
\label{tab:simple_knn_lr}
\begin{center}
{\renewcommand{\arraystretch}{1.2}\begin{tabular}{ c | c c c | c c c  }
\toprule
\multicolumn{1}{c|}{} & \multicolumn{3}{c|}{\tsg} & \multicolumn{3}{c}{\birdwatch} \\ 
\midrule
& \multicolumn{2}{|c }{kNN} & LR & \multicolumn{2}{|c}{kNN} & LR \\
 & $k=5$ & $k=10$ && $k=5$ & $k=10$ & \\ \hline
SiNE & 86.0 & 86.6 & 61.1  & 86.4 & 80.6 & 76.8  \\
StEM & 91.1 & 91.2 & 84.5  & 90.7 & 88.0 & 87.7  \\
SIDE & 91.0 & 87.5 & 82.1  & 92.6 & 90.0 & 82.7 \\
\midrule
SEM-mask & 90.5 & 92.3 & 84.4 & 92.4 & 90.4 & 86.6 \\
\midrule
SEM-addition & \textbf{94.0} & \textbf{95.3} & \textbf{88.1} & \textbf{94.6} & \textbf{92.9} & \textbf{91.5} \\
SEM-hadamard & 91.4 & 92.7  & 83.8 & 94.1 & 92.3 &  91.3\\

\bottomrule
\end{tabular}}
\end{center}
\end{table}

On \tsg~and \birdwatch, \sem-\textit{mask}, the topic-agnostic version of ~\sem, shows better/competitive performance with the three baselines. The topic-aware \sem~variants significantly outperform topic-agnostic baselines on both datasets and across both classifiers. On \tsg, \sem-addition improves the AUC by 2.9\% and 3.0\% the AUC for the $k=5$ and $k=10$ kNN classifiers respectively, compared to the best performing topic-agnostic method. On \birdwatch, \sem-addition improves the AUC by 2.0\% and 2.5\% the AUC at $k=5$ and $k=10$ respectively, compared to the best performing topic-agnostic method. These results demonstrate that \sem~learns improved node embeddings for signed edge prediction.

\subsection{Cold-start topic-stance detection} \label{subsec:zero-shot-stance-detection}

One important advantage of learning user and topic embeddings jointly is the potential for predicting the stance of a user on topics for which we have not observed their engagement. We investigate the performance of methods on this `cold start' subset of test samples $(u_1, u_2, w, t)$ such that the engagement of user $u_1$ or $u_2$ on topic $t$ was not observed during training. In other words, there is no training sample $(u_1, ., ., t)$ or $(., u_2, ., t)$. This represents 28\% and 17\% of the test data for \tsg~and \birdwatch~respectively (average over 5 folds). 
\begin{table}[h]
\caption{Mean AUC from 5-fold CV on cold-start stance detection using nearest neighbors (kNN). Results remain comparable to Table~\ref{tab:simple_knn_lr}, demonstrating we can effectively still maintain high accuracy without prior data on a user's interactions with a topic (Section~\ref{subsec:zero-shot-stance-detection}).} 
\label{tab:zssd}
\begin{center}
{\renewcommand{\arraystretch}{1.2}\begin{tabular}{ c | c c  | c c   }
\toprule
\multicolumn{1}{c|}{} & \multicolumn{2}{c|}{\tsg} & \multicolumn{2}{c}{\birdwatch} \\ 
\midrule
& \multicolumn{2}{|c }{kNN}  & \multicolumn{2}{|c}{kNN}  \\
 & $k=5$ & $k=10$ & $k=5$ & $k=10$  \\ \hline
SiNE & 83.0 & 84.3  & 84.2 & 80.1  \\
StEM & 92.8 & 90.2  & 89.9 & 88.4  \\
SIDE & 88.7 & 86.0  & 91.3 & 89.0\\
\midrule
SEM-mask & 87.9 & 90.0  & 90.5 & 90.1  \\
\midrule
SEM-addition & \textbf{95.1} & \textbf{96.1}  & \textbf{95.7} & \textbf{93.9} \\
SEM-hadamard & 90.4 & 90.1 & 93.4 & 92.4  \\
\bottomrule
\end{tabular}}
\end{center}
\end{table}

In Table~\ref{tab:zssd}, we present signed edge prediction AUC results limited to only `cold start' using only nearest neighbors classification since this had better performance overall. Only \sem-addition is able to maintain performance across both datasets and edge embedding classifiers (compared to Table~\ref{tab:simple_knn_lr}). This hints that, during training, \sem-addition learns topic relationships such that an observed disagreement on one topic affect the likelihood of disagreements (or agreements) for other topics.

\subsection{Learning topic embeddings for topic-agnostic approaches} \label{subsec:training_topic_embeddings}
We investigate learning topic embeddings separately from user node embeddings for topic-agnostic baselines. Because these methods do not jointly optimize the topic embeddings as part of a shared objective, we need to learn topic embeddings post- user embedding training. In order to do so, we alter how we train a link prediction classifier for topic-agnostic approaches to also learn a topic embedding table. For topic-aware \sem-methods, we instead opt to freeze this topic embedding table to what was learned during graph embedding. The intent of this experiment is to evaluate the value of jointly learning the topic embeddings along with node embeddings (as \sem~does), versus learning user and topic embeddings sequentially with methods agnostic to topics. \begin{figure}[h]
  \centering
  \includegraphics[width=.9\linewidth]{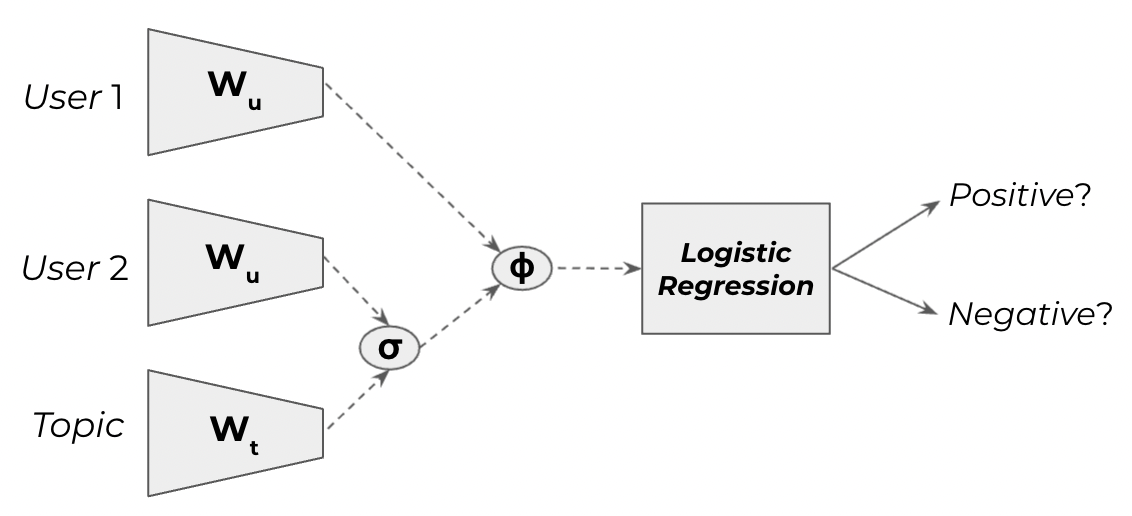}
  \caption{Logistic regression classifier for stance detection to investigate learning topic embeddings separately from user node embeddings (Section~\ref{subsec:training_topic_embeddings}).}
  \Description{A logistic regression classifier.}
  \label{fig:lr_classifier}
\end{figure}
As depicted in Figure \ref{fig:lr_classifier}, for a given edge $e=(u_1, u_2, w, t)$, this classifier takes as input the pre-trained user embeddings $u_1$ and $u_2$ combined with a topic embedding $t$ learned as part of this classifier training process for topic-agnostic approaches. We combine these embeddings similarly to how we propose in Section~\ref{subsec:overview} for training \sem: The user embedding $u_2$  and topic embeddings $t$ are combined via functions $\sigma(\cdot, \cdot)$ matching the choices for $\sigma$ that combine the graph-embedding learned user and topic embeddings defined in Section \ref{models}. The resulting vector is combined with $u_1$ user embedding vector via functions $\Phi(\cdot, \cdot)$ defined in Table~\ref{tab:operations}. The resulting edge embedding is the input to the LR classifier. Note that we deliberately combine the topic embedding with the user embedding $u_2$ only. Indeed edge operations $\ell_1$ and $\ell_2$ in Table~\ref{tab:operations} involve the difference between source and target node embeddings. So combining the topic embedding into source and target embeddings would cancel each other out. Note also that when we set $\sigma$=mask, we effectively ignore this learned (or frozen topic embedding), reducing to the same setting for LR in Table~\ref{tab:simple_knn_lr}. For other values of $\sigma$ the topic embedding (learned or frozen from graph embedding) is used for edge prediction. 

In Table \ref{tab:lrclassifier}, for each $\sigma$ and graph embedding approach, we report the best AUC found over functions $\Phi$. The performance of \sem-\textit{addition} remains unmatched by the topic-agnostic methods even when topic-agnostic approaches learn topic embeddings during classifier training. Performance is still significantly degraded compared to our best results in Table~\ref{tab:simple_knn_lr},  demonstrating that training topic and node embeddings in tandem remains the most beneficial way to incorporate context (topic) into stance detection on signed graphs. We do note however that for \sem-variant performance decreases if we use the learned topic embedding at test time.

\begin{table}[h]
\caption{Mean AUC from 5-fold CV on stance detection where we learn topic embeddings learned during link prediction, separately from graph embedding (Section~\ref{subsec:training_topic_embeddings}).} 
\label{tab:lrclassifier}
\begin{center}
{\renewcommand{\arraystretch}{1.2}\begin{tabular}{ c | c  c  c | c  c c }
\toprule
\multicolumn{1}{c|}{} & \multicolumn{3}{c|}{\tsg} & \multicolumn{3}{c}{\birdwatch} \\ 
\midrule
 & $\sigma=$ mask & add. & had. & mask & add. & had. \\ \hline
SiNE & 61.1 & 62.0 & 65.3 & 76.8 & 77.3 & 76.8 \\
StEM & 84.5 & 84.6 & 80.2 & 87.7 & 88.1 & 81.5 \\
SIDE & 82.1 & 82.2 & 81.1 & 82.7 & 82.6 & 79.5\\
\midrule
SEM-mask & 84.4 & 84.3 & 80.1 & 86.6 & 86.3 & 82.3\\
\midrule
SEM-addition & \textbf{88.1} & \textbf{88.1} & 81.2 & \textbf{91.5} & 89.8 & 86.9 \\
SEM-hadamard & 83.8 & 78.7 & 87.4 & \textbf{91.3} & 88.4 & 86.9 \\
\bottomrule
\end{tabular}}
\end{center}
\end{table}

\subsection{Visualising stance embeddings} \label{subsec:visualisation}

\begin{figure}[h]
  \centering
  \includegraphics[width=0.9\linewidth]{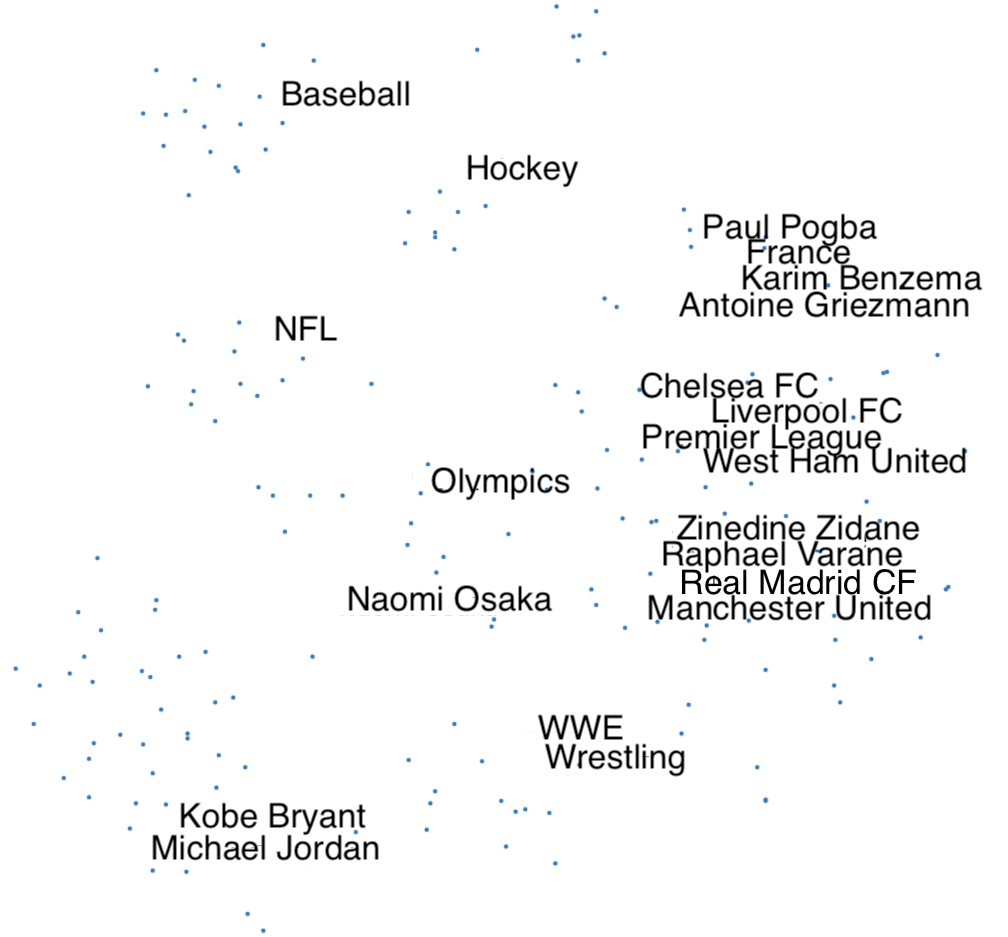}
  \caption{\tsg~topic embeddings learnt by \sem. Related topics are embedded close to each other. Only certain representative topics are labelled for readability.}
  \Description{A 2D-projection of topic embeddings.}
  \label{fig:tsg_topics}
\end{figure}

In Figure \ref{fig:tsg_topics}, we depict the topic embeddings obtained with \sem-\textit{addition} trained on \tsg, and projected with tSNE~\cite{van2008visualizing}. We can discern clear clusters of topics associated to a specific sport (e.g. NFL, hockey, baseball) or group of sports (e.g. fighting sports: WWE, Wrestling). Among these clusters, we observe finer-resolution groups. For instance, English football clubs lie close to the Premier League topic. Karim Benzema, Antoine Griezmann and Paul Pogba are the closest neighbours to France, while Zinedine Zidane and Raphael Varane are close to Real Madrid CF. Michael Jordan and Kobe Bryant are closest neighbours. We observe similar patterns on \birdwatch~topics, and with \sem-\textit{hadamard} (not depicted due to space constraints). The presence of meaningful topical clusters demonstrate the ability of our method to capture topic similarities when a diverse range of topics are discussed.

\begin{figure}[ht]
  \centering
  \includegraphics[width=\linewidth]{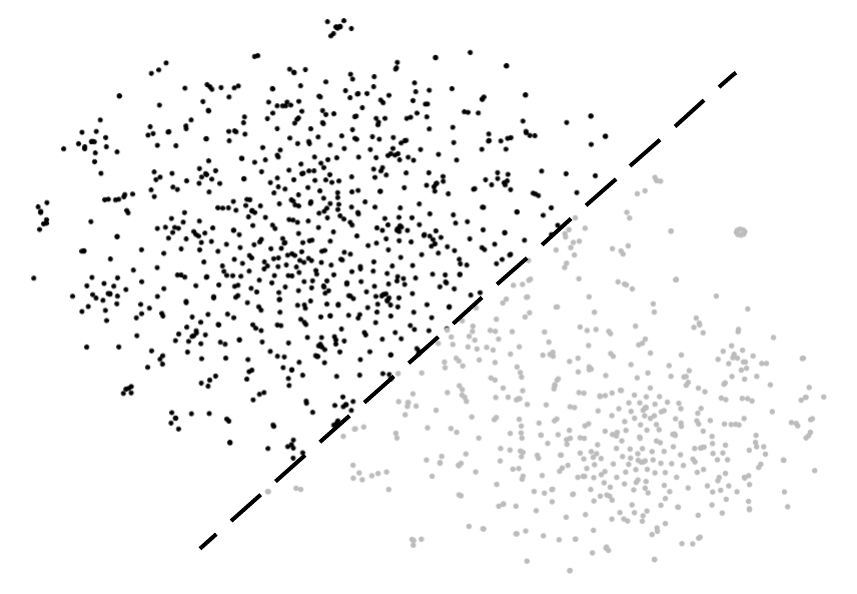}
  \caption{\birdwatch~user embeddings learnt by \sem. Two opinion communities are observed in Birdwatch.}
  \Description{A 2D-projection of user embeddings.}
  \label{fig:bsg_users}
\end{figure}

\balance

\begin{figure}[h]
  \centering
  \includegraphics[width=\linewidth]{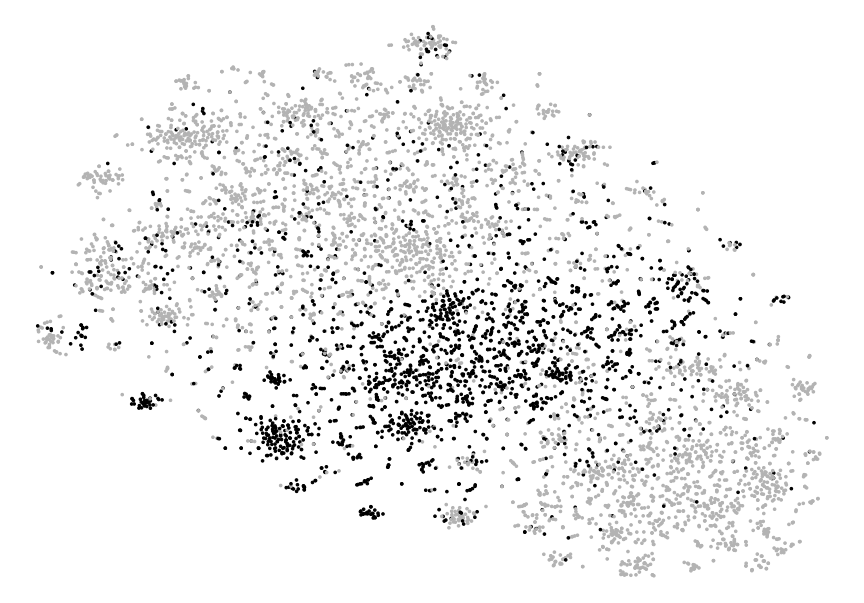}
  \caption{\birdwatch~edge embeddings derived from \sem's node embeddings. Negative edges are depicted in black, positive in grey appear to form distinct clusters.}
  \Description{A 2D-projection of edge embeddings.}
  \label{fig:bsg_edges}
\end{figure}

The US public debate and Birdwatch reports have been shown fall into two political party clusters~\cite{yasseri2021can, prollochs2021community, huang2021pole}. Consequently, as seen in Figure \ref{fig:bsg_users} which displays the user embeddings obtained with \sem-\textit{addition} trained on \birdwatch, and projected with tSNE, we observe two distinct opinion clusters -- verifying our sign-informed context generation strategy's ability to capture oppositions, and separate opposing views in the graph. Further, we visually inspect the ability of the model to distinguish positive and negative edges. Let $e=(u_1, u_2, w, t)$ be an edge of topic $t$ going from user $u_1$ to $u_2$ with weight $w \in \{-1, 1\}$. For visualisation, we define the embedding of edge $e$ as the hadamard product of the two user embeddings $u_1 \times u_2$. Figure \ref{fig:bsg_edges} displays the projected \birdwatch~edge embeddings obtained with \sem-\textit{addition} and tSNE. The positive (negative) edges are colored in dark (light). We observe distinct clusters of positive or negative edges, which confirms the capability of the model to discriminate positive and negative edges. 

\section{Conclusions}
In this work, we introduce \sem, a method for learning stance embeddings in signed, edge-attributed, social networks. Utilizing sign-informed random walks to generate training examples, we demonstrate how the scalable skip-gram objective can be successfully applied to learn contextualized signed-graph embeddings. Our approach is flexible and can incorporate arbitrary edge-attribute such as topics, to exploit contextualised edges and unveil insights about the users and the graph attributes. Experimental results show that \sem ~embeddings outperform state-of-the-art signed-graph embedding techniques on two new Twitter datasets: \tsg~and  \birdwatch. We open-source these two datasets to the network mining community to spur further research in social network analysis and stance detection.

\color{black}

\begin{acks}
We thank Renaud Lambiotte for his valuable feedback.
\end{acks}

\bibliographystyle{ACM-Reference-Format}

\bibliography{stance_wsdm}

\end{document}